\documentclass[11pt,a4paper]{article}
\usepackage{times,latexsym}
\usepackage{url}
\usepackage[T1]{fontenc}

\usepackage[]{emnlp2023}
\usepackage{xspace,mfirstuc,tabulary}

\usepackage{booktabs}
\usepackage{subfigure}
\usepackage{float}
\usepackage{makecell}
\usepackage{hyperref}
\usepackage{graphicx}
\usepackage{booktabs}
\usepackage{arydshln}
\usepackage{tabularx}
\usepackage{enumitem}
\usepackage{subfigure}
\usepackage{multirow}
\usepackage{amsmath}
\usepackage{bm}
\usepackage{amsfonts}
\usepackage{pgfplots}
\pgfplotsset{compat=1.7}
\usepackage{pifont}
\usepackage{colortbl}
\usepackage{algorithm}
\usepackage{algpseudocode}

\usepackage{CJK}

\newif\iftaclinstructions
\taclinstructionsfalse % AUTHORS: do NOT set this to true
\iftaclinstructions
\renewcommand{\confidential}{}
\renewcommand{\anonsubtext}{(No author info supplied here, for consistency with
TACL-submission anonymization requirements)}
\newcommand{\instr}
\fi

\title{TransCoder: Towards Unified Transferable Code Representation Learning Inspired by Human Skills}

\author{Qiushi Sun$^{1,2}$, Nuo Chen$^1$, Jianing Wang$^1$, Xiang Li$^{1}$\thanks{ \quad Corresponding Author.}, Ming Gao$^{1}$\\
  $^1$ East China Normal University, Shanghai, China \\
  $^2$ National University of Singapore, Singapore\\
 \texttt{\{qiushisun,nuochen\}@stu.ecnu.edu.cn,} \texttt{lygwjn@gmail.com} \\
  \texttt{\{xiangli,mgao\}@dase.ecnu.edu.cn}
}

\date{}

\begin{document}
\maketitle

\begin{abstract}
Code pre-trained models (CodePTMs) have recently demonstrated a solid capacity to process various software intelligence tasks, 
\textit{e.g.}, 
code clone detection, code translation, and code summarization.
The current mainstream method that deploys these models to downstream tasks is to 
fine-tune them on individual tasks,
which is 
% However,
generally costly
and 
needs sufficient data 
for large models.
To tackle the issue,
in this paper, 
we present \textbf{TransCoder}, 
a unified \underline{Trans}ferable fine-tuning strategy for \underline{Code} \underline{r}epresentation learning.
Inspired by human inherent skills of knowledge generalization,
%and cross-domain transfer
TransCoder drives the model to learn better code-related knowledge like human programmers.
Specifically, 
we employ a tunable prefix encoder to first capture cross-task and cross-language transferable knowledge, 
and then apply the acquired knowledge for downstream adaptation. 
Besides, 
tasks with minor training sample sizes and languages with small corpus can be remarkably benefited from our approach.
Extensive experiments conducted on representative datasets clearly demonstrate that 
our method can lead to
superior performance on various code-related tasks and encourage mutual reinforcement.
We also show that TransCoder is applicable in low-resource scenarios. 
Our codes are available at \url{https://github.com/QiushiSun/TransCoder}.

% \footnote{All codes and datasets will be released upon acceptance.}.
\end{abstract}

% Introduction
\section{Introduction}
With the remarkable success pre-trained language models~\cite[][\emph{inter alia}]{devlin2018bert,radford2019language,liu2019roberta,raffel2019exploring, qiu2020pre} have achieved in natural language processing (NLP), 
leveraging pre-training strategies has gradually become the de-facto paradigm for neural code intelligence.
Under the assumption of ``Software Naturalness''~\cite{HindleBGS16}, researchers have proposed several code pre-trained models (CodePTMs)~\cite{ feng2020codebert, guo2022unixcoder,xu2022survey,xu2022systematic}, 
which utilize unsupervised objectives in the pre-training stage 
and are fine-tuned on downstream tasks.
These CodePTMs have shown great capacity in code understanding~\cite{mou2016convolutional, svajlenko2014towards} and code generation~\cite{nguyen2015divide, husain2019codesearchnet}.
The applications of code intelligence in real-world scenarios can be mainly divided into two dimensions: languages and tasks; 
when adapting these powerful CodePTMs to downstream tasks of code representation learning, 
each model requires to be tuned on a specific task in a particular language. 
This faces challenges that include data insufficiency and imbalance.
% \sqs{but... these models are usually tuned on specific tasks..., problems on data imbalance, data insufficiency, training cost need attention}
% maml -> GOOD INIT
% meta-learning -> quickly adapt to downstream
\begin{figure*}[ht!]
    \centering
    \subfigure[Cross-task code representation learning]{ % width=3.6cm
        \includegraphics[width=0.53\textwidth]{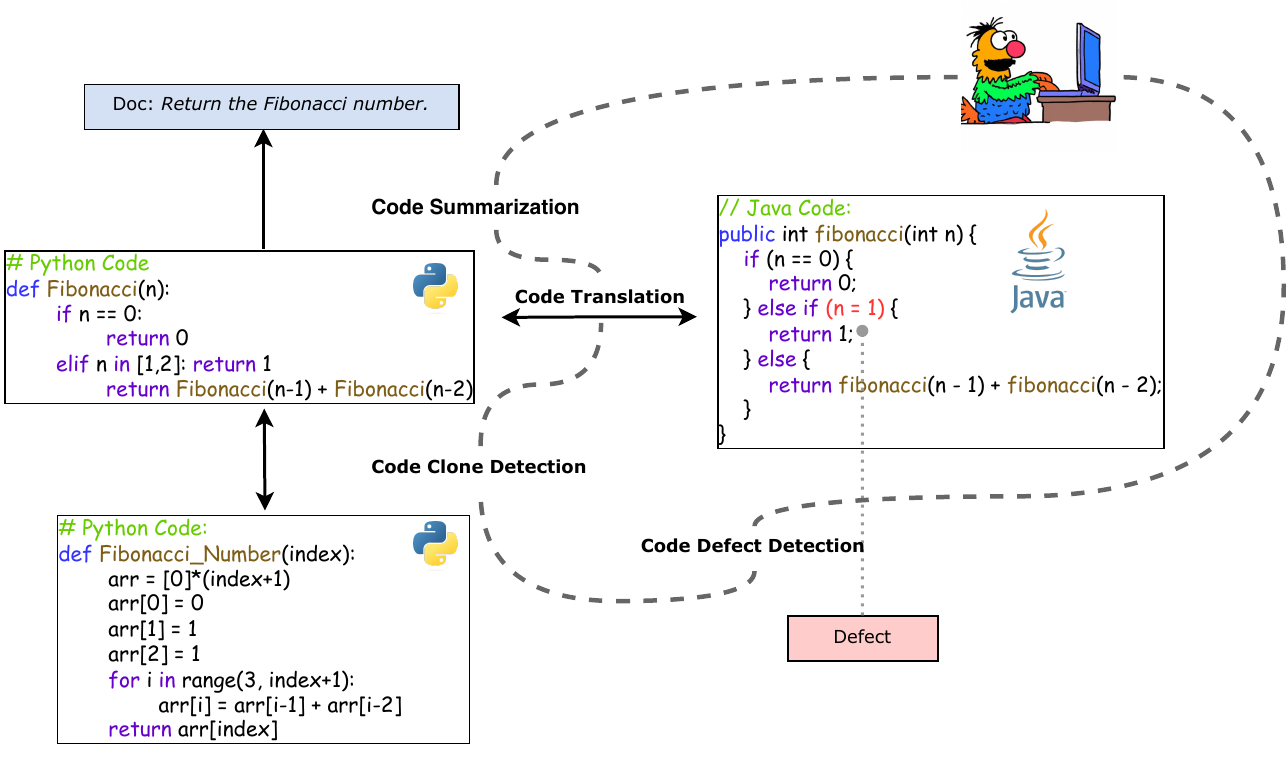}
        \label{fig:cross-task}
    }
    \subfigure[Cross-language code representation learning]{
	\includegraphics[width=0.43\textwidth]{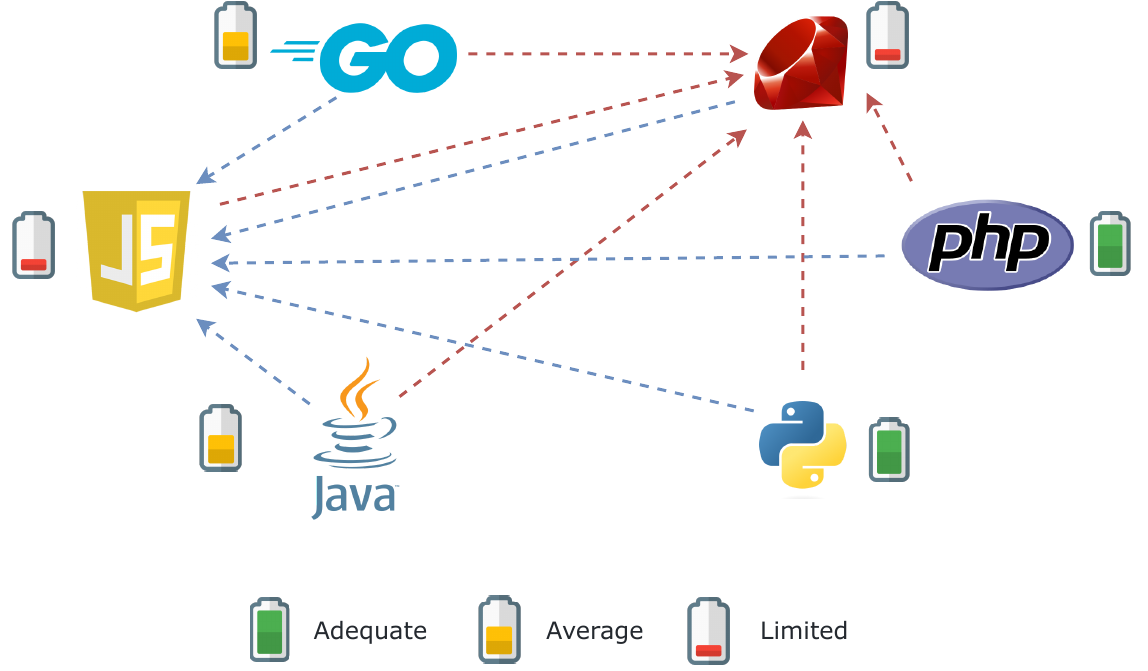}
	\label{fig:cross-lang}
    }
    \caption{(a) A CodePTM~(\textit{e.g.}, CodeT5, PLBART) will learn through a series of code downstream tasks such as code summarization and clone detection in the learning process, 
    in order to acquire cross-task knowledge of code representation.
    (b) In the currently available code corpora (both bimodal and unimodal data), 
    there is an imbalance between different PLs.
    Nonetheless, different languages share similar programming principles so that they can ``support'' each other through the models' learning cross-language knowledge.
    (Best viewed in color.)}
    \label{toktype}
    \vspace{-0.75em}
\end{figure*}

% from Human Intelligence

To address this kind of problem, the idea of meta-learning has been widely adopted for model adaptation~\cite{finn2017maml, jamal2019taml, rajasegaran2020itaml}. More recently, \citet{wang2021transprompt} employs a multitask meta-knowledge acquisition strategy to train a meta-learner that can capture cross-task transferable knowledge from a series of NLP tasks.
However, 
learning the transferable representation of code is under a more complex scenario in the ``Big Code'' era~\cite{allamanis2018survey}, as multiple task types should be tackled, 
and a diversity of programming languages (PLs) must also be considered. Thus, applying previous approaches in NLP is less feasible, 
while utilizing the idea of a “meta-leaner” for knowledge transfer appears to be promising in the field of code intelligence.
Therefore, a research question naturally arises: \textit{Can we develop a unified learning framework for CodePTMs that captures transferable knowledge across both code-related tasks and programming languages to enhance the code representation learning? }

As a programmer,
the programming learning path of humans has told us that: 
the experience of performing multiple code-related tasks~(\textit{e.g.}, debugging, writing docs)
or learning various programming languages~(\textit{e.g.}, C, Python) should not cancel each other out but be able to reinforce each other. 
% \sqs{Learning multiple code-related tasks or programming languages should not cancel each other out but be able to reinforce each other. +more details}
By understanding the fundamentals of the first programming language, learning another one that utilizes similar principles becomes more straightforward. 
% As the number of mastered languages increases for a programmer, 
% it becomes more accessible to learn the next one. 
As a programmer masters more PLs, it becomes progressively attainable to pick up the next one.
These are the innate human skills of knowledge generalization.
% and refinement.

Inspired by the aforementioned human skills~\cite{zhu2022neural},
this work proposes TransCoder, a unified transferable code representation learning framework.
In particular,
% As is shown in
Figure~\ref{fig:cross-task}
% which 
showcases the procedure of learning multiple code-related tasks that cover both code understanding and generation.
Then,
figure~\ref{fig:cross-lang} illustrates the scenario of using other PLs with an abundant volume of data\footnote{the data includes bimodal data that refer to parallel data of NL-PL (Natural Language-Programming Langauge) pairs and unimodal stands for pure codes without NL comments.} 
to enhance the one with insufficient examples,
which is similar to multilingual learning in NLP~\cite{ustun2022hyper}.
We describe this procedure of learning cross-task and cross-language representation as \textit{universal code-related knowledge acquisition}.
To be specific,
a transferable prefix that plays the role of ``universal-learner'' is employed to acquire and bridge the knowledge of different tasks and languages based on continual learning~\cite{chenLML2018}. 
This process can be analogous to people learning different languages and using them in various contexts.
% When enough knowledge has been absorbed in a given task, 
Once a sufficient amount of knowledge has been ``assimilated'' for a particular task,
the prefix will be concatenated to a fresh CodePTM to commence the next phase of training, and so forth.
It is a similar process for cross-language knowledge acquisition by TransCoder.

In addition, 
the nature of transferable representations allows us to conduct few-shot learning on downstream tasks, 
which can serve as an alternative under extreme data scarcity.
Extensive experiments among several PLs and code-related tasks demonstrate the effectiveness of TransCoder. 
Our main contributions are summarized as follows:
% \item We ... mutual reinforcement
% \item Structural information
\begin{itemize}[itemsep=5pt,topsep=0pt,parsep=0pt]
    \item In this paper, we propose TransCoder, which is a novel transferable framework 
    for both code understanding and generation.
    \item A prefix-based universal learner is presented to capture general code-related knowledge, enabling unified transferability and generalization.
    Besides, 
    TransCoder makes it feasible to drive cross-language code representation learning 
    that empowers PLs with insufficient training data to conquer complex tasks.
    \item Extensive experiments
    demonstrate the capability of TransCoder, 
    which not only boosts reciprocal reinforcement among tasks but also mitigates the unbalanced data problem.
\end{itemize}

% Background & Related works
\section{Related Works}

In this section, 
we first briefly summarize the related works on code pre-trained models. 
Then we introduce the idea of transfer learning and continual learning.

\paragraph{Pre-trained Language Models for Code}

Pre-trained language models have significantly advanced performance in a broad spectrum of NLP tasks.
Inspired by the success of Transformer-based pre-trained models, 
attempts to apply pre-training to source code have emerged in recent years~\cite{xu2022systematic, zan2023large}.
\citet{feng2020codebert} first propose CodeBERT, 
which is pre-trained on NL-PL pairs in six languages with masked language modeling (MLM) and replaced token detection (RTD)~\citep{yang2019xlnet} objectives. 
GraphCodeBERT~\cite{guo2021graphcodebert} first utilizes the information contained in Abstract Syntax Tree (AST), and extracts code structures from source code to enhance pre-training. 
PLBART~\cite{ahmad2021unified} adapts the BART architecture and is pre-trained with denoising objectives on Python/Java code data and NL posts. 
CodeT5~\cite{wang2021codet5,wang2023codet5plus} builds on the T5~\cite{raffel2019exploring} model architecture that supports both code understanding and generation. 
It allows for multi-task learning and considers crucial token types (\textit{e.g.}, identifiers). 
Recently proposed UniXcoder~\cite{guo2022unixcoder} uses a multi-layer Transformer-based model that follows UniLM~\cite{dong2019unified} to utilize mask attention matrices with prefix adapters.

\paragraph{Transfer-Learning}
Over the past few years, 
there have been substantial advancements in transfer-learning techniques within the NLP community, 
leading to significant improvements in a wide range of tasks~\cite{ruder2019transfer}.
% In recent years, 
% the NLP community has witnessed considerable advancement in several transfer-learning methods that significantly improved upon the state-of-the-art on a broad scope of NLP tasks~\cite{ruder2019transfer}.
Transfer-learning leverages the knowledge contained in related source domains to enhance target learners' performance on target domains~\cite{zhuang2021transfer}. 
Previous research demonstrates effective transfer from data-rich source tasks~\cite{phang2018sentence}, 
especially for the scenario that reasoning and inference are required~\cite{pruksachatkun2020intermediate}.
When considering various tasks as related, 
transfer learning among multitasks has already shown superior performance compared to single-task learning~\cite{aghajanyan2021muppet}.
\citet{wang2021transprompt} employ prompt tuning-based methods to transfer knowledge across similar NLP tasks for the purpose of mutual reinforcement.
Recently,~\citet{vu2022spot} adopt the method of learning a prompt on one or more source tasks and utilizing it to initialize the prompt for target tasks.

%  knowledge can be transferred across similar NLP tasks for the purpose of mutual reinforcement
% prompt tuning-based ideas to transfer learning

\paragraph{Continual Learning}
Continual learning is also referred to as incremental or lifelong learning~\cite{chenLML2018, rao2019continual, german2019continual}, which is inspired by the learning process of human beings.
The primary criterion is acquiring knowledge gradually 
while preventing catastrophic forgetting~\cite{Kirkpatrick2017overcoming}
and then applying the knowledge for future learning.
To be specific, 
the model undergoes training on multiple tasks or datasets in a sequential manner
% is trained with several tasks/a continuous flow of data sequentially 
in order that remembers the previous tasks when learning the new ones.
Recent studies on continual learning propose combinations of different techniques~\cite{cao2021continual} to reduce the risk of forgetting
and continual pre-training~\cite{Sun2020ernie} 
that supports customized training tasks.

% In contrast, 
% meta-learning is about 'learning to learn', \textit{i.e.}, seeks to learn models that can quickly adapt to other tasks or environments with a few training examples~\cite{wang2020generalizing,huisman2021deepmeta}. 
% Meta-learning algorithms have already covered a series of typical NLP tasks~\cite{lee2022meta}, such as text classification~\cite{geng2020dynamic}, sequence labeling~\cite{hou2020shot} and knowledge completion~\cite{sheng2020adaptive}.
% Different meta-learning methods focus on learning various elements for more generalizable models or faster convergence. 
% For instance, MAML (model-agnostic meta-learning)~\citep{finn2017maml, antoniou2018how} focuses on learning better parameter initialization. 

% and meta fine-tuning~\cite{wang2020meta} to solve similar NLP tasks.

% Model
\begin{figure*}[t]
    \centering
    \includegraphics[width=\linewidth]{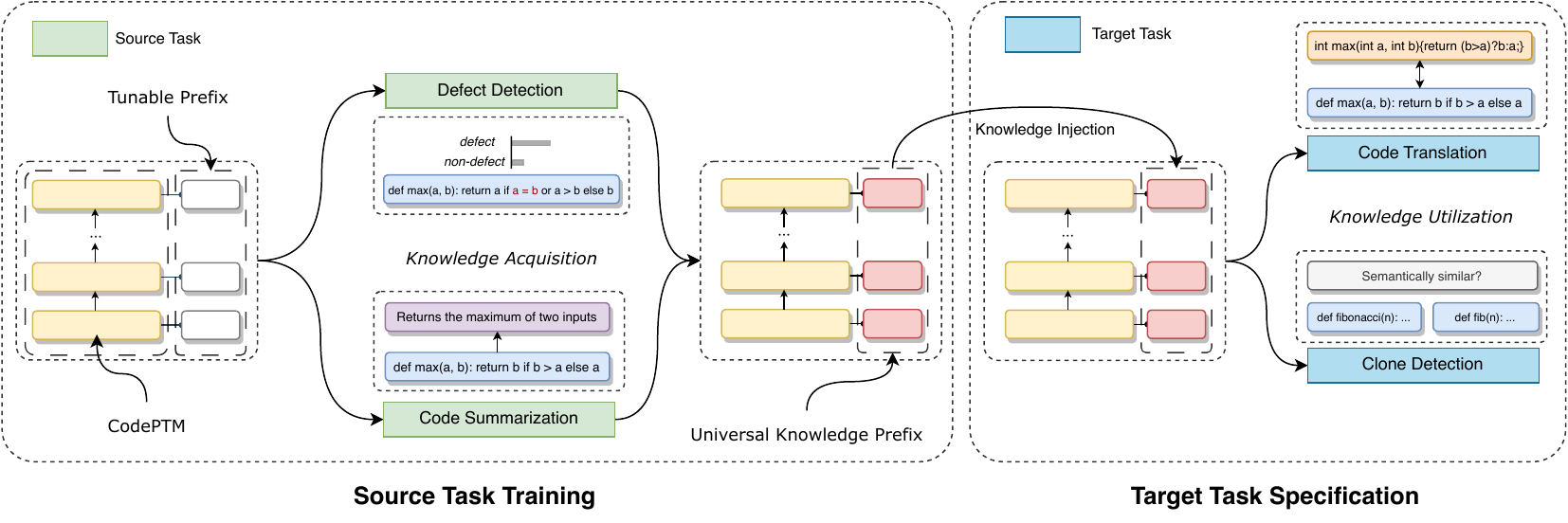}
    \caption{An illustration of the architecture of TransCoder.
    (1) In the source task training stage, tunable universal knowledge prefixes are first randomly initialized and prepended with a CodePTM~(\textit{e.g.}, CodeT5, PLBART). The whole model is tuned by back-propagation.
    (2) For the target tasks specification stage, 
    we prepend these universal knowledge prefixes to a new CodePTM,
    effectively infusing universal knowledge into the model.
    For brevity, 
    we choose a cross-task scenario and use some representative tasks as illustrations in this figure, 
    which means using the knowledge acquired from code summarization/defect detection to enhance the performance of code translation/clone detection.
    The order of tasks or languages could be rearranged flexibly.
    (Best viewed in color.)}
    \label{fig:TransCoder-Arch}
    \vspace{-1em}
\end{figure*}

% In the target task specification stage, we prepend these universal knowledge prefixes to a new CodePTM, effectively infusing universal knowledge into the model.

\section{TransCoder}

In this section, 
we formally present the model architecture and the universal knowledge acquisition process of TransCoder.\footnote{
For a more precise understanding, 
here we clarify that TransCoder acquires cross-task and cross-language knowledge by employing continual learning.
Our core idea is to learn universal knowledge 
across tasks, 
not to learn a better parameter initialization oriented toward few-shot learning or fast adaptation.
Even so, 
TransCoder can indeed support low-resource scenarios, as is shown in section~\ref{method:low-resouce}.}

\subsection{Overview}

\citet{bengio2021deep} indicates the next stage of deep learning by comparing current AI approaches with the learning capabilities of human beings. 
In the domain of code intelligence,
a recent work~\cite{zhu2022neural} considers human behaviors in reading programs. 
In this paper, 
we take a big step forward in that 
TransCoder takes human \textit{inherent skills} into account.
Like human programmers, 
the models have the capacity to learn programming languages incrementally 
and enhance their overall comprehension of codes when being exposed to related tasks.
Specifically, we define the transferable knowledge of code representation learning as \textit{universal code-related knowledge}.
As is demonstrated in Figure~\ref{fig:TransCoder-Arch},
\textit{knowledge prefixes} concatenated with the CodePTMs play the role of the \textit{universal knowledge provider} that bridges different downstream tasks. 
The first stage of TransCoder is \textit{Source Tasks Training}, acquiring cross-domain and cross-language knowledge by continual learning~\cite{Sun2020ernie}, which is the core idea that enables CodePTM to ``learn coding'' like humans. 
The second stage is \textit{Target Task Specification}, which applies the learned knowledge to new unseen code-related tasks through prefix concatenation.

% \sqs{Method -> unified}
% \subsection{Notation and Definition}
% There are four representative code downstream tasks involved \sqs{break translate and summarization}
% \subsubsection{Definition}
% \vspace{-.1em}
% \paragraph{Universal Knowledge} Learning the universal knowledge $\theta$ from various downstream tasks across different programming languages is the core idea of TransCoder. \sqs{add details}

\subsection{Universal Knowledge Prefix}
\label{sec:kprefix}
% \sqs{this section add details, see transprompt}
Inspired by deep prompt-tuning~\cite{li2021prefix, liu2022ptuning, xie2022unifiedSKG}, 
which prepends tunable prefixes to hidden states at every layer of the transformer-based models.
% we propose a  prefix to conduct universal knowledge acquisition and transfer.
We employ knowledge prefixes $\mathcal{F}_{uni}$ that could be used to acquire universal code-related knowledge. 
The knowledge prefixes are then concatenated to the CodePTM $\psi_i$ for both source task training and target task specification.
Precisely, the prefixes inject universal knowledge into the attention modules of the CodePTMs,
both the prefix and the knowledge injection procedure would not be constrained by the model backbone~(\textit{e.g.}, CodeT5, PLBART).
Moreover, the knowledge prefix further supports cross-task and cross-language learning in a unified way: 
No matter what kinds of tasks we are tackling or how many languages are involved,
the usage and effectiveness remain the same.

\subsection{Training Procedure}

\paragraph{Source Tasks Training.}
The first stage of TransCoder is source task training,
which aims to acquire and transfer code-related knowledge to the universal knowledge prefix.
We define a task that provides universal code-related knowledge as a source task\footnote{For the code summarization task, the training data for each source task corresponds to a sub-dataset of CodeSearchNet dataset partitioned by different programming languages.} $t$, and the collection of $k$ source tasks as $T$. 
For code summarization that has six different PLs, we "break" it into six independent source tasks. Likely, the task of code translation between C\# and Java is defined as two source tasks as well. 
We note their training data as source data $D(t)$. 
Figure~\ref{fig:cross-task} gives an intuitive understanding of source task training.
Similarly, 
Figure~\ref{fig:cross-lang} provides a straightforward illustration of using tasks in other programming languages for universal knowledge acquisition and then supporting target tasks.

% \begin{algorithm}[ht]
% 	\caption{Source Tasks Training}
% 	\begin{algorithmic}[1]
% 		\Require \
%             		 $t$, $\mathcal{F}_{uni}$, $\theta$, $f_t$ , $\psi_i$
% 		\Ensure \
% 		Universal knowledge: $\theta$
% 		\Function % $t\_list$
% 		{$Trainer$}{$\mathcal{F}_{uni}$,$\theta$,$f_t$,$\psi_i$}  
% 		\State Randomized $\theta$
%             \For{$e$ in epochs}
%             \For{$i$ in tasks}
% 		\State Adaptive Sample from $D(i)$
%             \State $loss_i \leftarrow L_{code}(f_i(\mathcal{F}_{uni}(\theta),\psi_i))$
%             \State $\theta,\psi_i \leftarrow \theta,\psi_i - \alpha \nabla_{\theta}loss_i(\theta,\psi_i)$
%             \EndFor
%             \EndFor
% 	    \EndFunction
% 	\end{algorithmic}
%  \label{alg:main}
% \end{algorithm}

\begin{algorithm}[ht]
	\caption{Source Tasks Training}
	\begin{algorithmic}[1]
		\Require 
            Source task $t$, 
            Universal knowledge prefix $\mathcal{F}_{uni}$, 
            Universal knowledge $\theta$, 
            Task-specific model $f_t$ , 
            CodePTM $\psi_i$
		\Ensure 
		Updated universal knowledge: $\theta$
		
		\Function{$Trainer$}{$\mathcal{F}_{uni}$,$\theta$,$f_t$,$\psi_i$}  
		\State Randomly initialize $\theta$
        \For{each epoch $e$}
            \For{each task $i$ in $t$}
		    \State Adaptive sample data batch $D(i)$ from source tasks.
            \State Get $loss_i\leftarrow L(f_i(\mathcal{F}_{uni}(\theta),\psi_i))$
            \State Update universal knowledge and CodePTM model parameters: 
            \Statex \hspace{2em} $\theta, \psi_i \leftarrow \theta, \psi_i - \alpha \nabla_{\theta,\psi_i} loss_i$
            \EndFor
        \EndFor
	    \EndFunction
	\end{algorithmic}
 \label{alg:main}
\end{algorithm}
\vspace{-0.5em}

% \begin{algorithm}[ht]
% 	\caption{Source Tasks Training}
% 	\begin{algorithmic}[1]
% 		\Require \
% 		 % $t$: task, Data for task $t$:$\mathcal{D}(t)$ \\
%    %          prefix: $\mathcal{F}_{meta}$ \\
%    %          knowledge: $\theta$ \\
%    %          backbone CodePTM model: $f_t$ \\
%    %          model param: $\psi_t$
%             		 $t$, $\mathcal{F}_{uni}$, $\theta$, $f_t$ , $\psi_i$
% 		\Ensure \
% 		% A flatted sequence tokens $seq$ %that can represent the AST
% 		Universal knowledge: $\theta$
% 		\Function % $t\_list$
% 		{$Trainer$}{$\mathcal{F}_{uni}$,$\theta$,$f_t$,$\psi_i$}  
% 		\State Randomized $\theta$
%             \For{$e$ in epochs}
%             \For{$i$ in tasks}
% 		\State Adaptive Sample from $D(i)$
%             % \For{$b$ in $B$}
%             \State $loss_i \leftarrow L_{code}(f_i(\mathcal{F}_{uni}(\theta),\psi_i))$
%             \State $\theta,\psi_i \leftarrow \theta,\psi_i - \alpha \nabla_{\theta}loss_i(\theta,\psi_i)$
%             %\EndFor
%             \EndFor
%             \EndFor
% 	    \EndFunction
% 	\end{algorithmic}
%  \label{alg:main}
% \end{algorithm}
% \vspace{-0.5em}

\paragraph{Universal Knowledge Acquisition.}
% A brief description of how TransCoder acquires universal knowledge related to code is provided in Algorithm~\ref{alg:main}.
Algorithm~\ref{alg:main} briefly describes how TransCoder acquires code-related universal knowledge. 
% Sample\footnote{The "Adaptive Sample" mentioned in Algorithm~\ref{alg:main} refers to the smoothed sampling strategy due to the imbalance of training data of each source task.}
The inner loop employs the strategy of continual learning~\cite{Sun2020ernie}.
For each task $t$ sampled from $T$, the trainer uses back-propagation to optimize the entire model with the cross-entropy loss.
In this process, both the parameters of the knowledge prefix and the CodePTM $\psi_i$ are updated by the back-propagation.
The outer loop ensures the knowledge prefix ``absorbs'' sufficient universal knowledge through $e$ epochs of source task training. In each epoch, the trainer will iterate through all of the source tasks.

In practice, 
we first randomly select a source task $t$ from available tasks and then sample a batch from the task's training data $\mathcal{D}(t)$. 
The knowledge prefix $\mathcal{F}_{uni}$ first holds random knowledge $\theta_{0}$ , \textit{i.e.}, initialized with random parameters.
During source task training:

% \sqs{for each epoch training.., as follow:}

$$\overbrace{\theta_{0} \longrightarrow \theta_{1}}^{T_{0}\cdots T_{k}} \rightarrow \cdots \rightarrow  \theta_{e-1} \rightarrow \theta_{e}$$

Each transition refers to the trainer iterating through all the $k$ source tasks in one epoch. 
During this procedure, the knowledge prefix $\mathcal{F}_{uni}$ will be moved to a new CodePTM whenever a new task comes, as is shown in Figure~\ref{fig:TransCoder-Arch}.
After $e$ epochs of training, $\mathcal{F}_{uni}$ captures enough universal code-related knowledge among different tasks or PLs.

% ...\sqs{train e round to ensure the acquisition of code-related knowledge}
% Alg~\sqs{due to the imbalance of data of each source task, we use a smoothing method}
\paragraph{Source Data Sampling.}

In source task training, 
we are aware that the training sample sizes of each source task might vary considerably. 
Learning directly from these datasets would make the universal knowledge prefix biased toward tasks with large datasets.
Thus, we employ an ``Adaptive Sampling'' strategy.
When TransCoder samples $M$ training data from $\mathcal{D}(1), \mathcal{D}(2), \cdots, \mathcal{D}(M)$, instead of randomly selecting training samples,
it employs stratified sampling where training instances are selected with the probability proportional to the dataset distribution $P(\mathcal{D}(k))$:
\begin{equation}
P\left(\mathcal{D}(k)\right)=\frac{\log \left|\mathcal{D}(k)\right|+\delta}{\sum_{\tilde{k}=1}^M \log |\mathcal{D}({\tilde{k})}|+\delta},
\label{smoothed-sampling}
\end{equation}
where $\delta$ > 0 is the smoothing factor. 
This strategy results in the over-sampling of small datasets and the under-sampling of large datasets.

\paragraph{Target Task Specification.}
% Target task specification 
This is the second stage of TransCoder.
As is detailed in Algorithm~\ref{alg:main},
the universal knowledge acquired in source task training grants us knowledge prefixes that could be concatenated to new CodePTMs. 
It enhances the target task's performance by injecting code-related knowledge $\theta_{e}$ by prefix $\mathcal{F}_{uni}$.
Figure~\ref{fig:TransCoder-Arch} gives an example of utilizing universal knowledge to enhance target tasks.
The CodePTM with the universal knowledge prefix is trained together in this stage for downstream adaptation.

\subsection{Learning under Low-resource Scenarios}
\label{method:low-resouce}
Apart from learning from the full dataset, 
the universal knowledge acquired from source tasks innately enables TransCoder to conduct code-related tasks under low-resource scenarios.

Under this setting, both the source tasks training algorithm and data sampling strategy are identical to Algorithm~\ref{alg:main} and Equation~\ref{smoothed-sampling} respectively.
However, in the target task specification stage, we only need to utilize 5\% - 20\% data for downstream adaption. The settings are detailed in section~\ref{exp:low-resource}.

% \input{tables/cross-task-main.tex}

% Experiments
\begin{table*}[t]
\centering
\resizebox{0.9\textwidth}{!}{
\begin{tabular}{lcccccccccccc}
\toprule
\multirow{3}{*}{\textbf{Methods}} & \multicolumn{2}{c}{CLS2Trans} & \multicolumn{2}{c}{Sum2Trans}  & \multicolumn{1}{c}{CLS2Sum}  & \multicolumn{1}{c}{Trans2Sum} & \multicolumn{2}{c}{Sum2CLS} & \multicolumn{2}{c}{Trans2CLS} \\ 
\cmidrule(lr){2-3}\cmidrule(lr){4-5}\cmidrule(lr){6-6}\cmidrule(lr){7-7}\cmidrule(lr){8-9}\cmidrule(lr){10-11}
& BLEU & EM & BLEU & EM & BLEU & BLEU & \makecell[c]{Clone \\ \small{F1}} & \makecell[c]{Defect \\ \small{Acc}} & \makecell[c]{Clone \\ \small{F1}} & \makecell[c]{Defect \\ \small{Acc}} \\

\midrule
\textbf{CodeT5} &  &  &  &  &  &  &  &  &  &  \\
Fine-Tuning & \textbf{81.63} & 65.80 & 81.63 & 65.80 & 19.56 & 19.56 & \textbf{94.97} & 64.35 & 94.97 & 64.35\\
% TransCoder\textsubscript{w/o} & *** & *** & *** & *** & *** & *** & *** & *** & *** & ***\\ \textsubscript{full}
TransCoder & 81.43 & \textbf{67.00} & \textbf{82.12} & \textbf{68.20} & \textbf{20.39} & \textbf{19.77} & 93.70 & \textbf{66.58} & \textbf{95.39} & \textbf{66.36} \\
\midrule
\textbf{PLBART} &  &  &  &  &  &  &  &  &  & \\
Fine-Tuning & \textbf{78.17} & 62.70 & \textbf{78.17} & \textbf{62.70} & 17.93 & 17.93 & \textbf{92.85} & 62.27 & 92.85 & 62.27\\
% TransCoder\textsubscript{w/o} & *** & *** & *** & *** & *** & *** & *** & *** & ***  & ***\\ \textsubscript{full}
TransCoder & 76.00 & \textbf{63.40} & 74.50 & 56.00 & \textbf{18.62} & \textbf{18.25} & 92.28 & \textbf{64.58} & \textbf{92.91}  & \textbf{64.98}\\
\bottomrule
    \end{tabular}
    }
    \caption{The performance on the code cross-task learning with CodeT5 and PLBART backbone.
    The code translation task between C\# and Java is evaluated on the averaged BLEU-4 scores~\cite{papineni2002bleu} and Exact Match (EM).
    For code summarization, 
    we report the average of TransCoder's performance with smoothed BLEU-4~\cite{lin2004orange} metrics on six programming language tasks.
    At last, two code understanding tasks: clone detection and defect detection are evaluated on F1 score and accuracy respectively.
    }
    \label{tab:cross-task}
    \vspace{-1.0em}
\end{table*}

\section{Experiments}

In this section, we demonstrate the performance of TransCoder on cross-task and cross-language code representation learning, 
including the scenario of learning from inadequate data.

% \vspace{-.1em}
\subsection{Tasks and Languages} We conduct our experiments on four code representation learning tasks from the CodeXGLUE benchmark~\cite{Lu2021codexglue} that can be divided into code understanding and code generation.

% \vspace{-.1em}
\paragraph{Code Generation.} For code generation tasks, code summarization~\cite{alon2018codeseq} aims to generate natural language comments for the given code snippet in a different programming language. We use the CodeSearchNet dataset for the experiment, and the statistics of data are listed in table~\ref{table:csn-stats}.
Additionally, we can observe a huge gap between the data volume of Java and Ruby.

The second code generation task is code translation~\cite{nguyen2015divide}. This task involves translating a code snippet from one programming language to another. Table~\ref{table:code-trans} demonstrate the statistics of the Java-C\# dataset used in the experiment.
Compared with the data volume of code summarization, the CodeTrans dataset contains far fewer training samples.

\paragraph{Code Understanding.} Code understanding tasks evaluate models' capability of understanding code semantics and their relationships. 
Clone detection~\cite{svajlenko2014towards, mou2016convolutional} measures the similarity between two code snippets. 
Defect detection~\cite{zhou2019devign} seeks to predict whether the source code contains vulnerability to software systems.
In our experiment, 
we use the BigCloneBench dataset in Java language and Devign dataset in C language, as is shown in table~\ref{table:clone-defect-stats}.
\vspace{-0.25em}

\subsection{Experimental Setup}

\paragraph{Backbone Models.} In our experiments, we mainly use two CodePTMs with encoder-decoder architecture: CodeT5-base\footnote{\href{https://huggingface.co/Salesforce/codet5-base}{huggingface.co/Salesforce/codet5-base}}~\cite{wang2021codet5} and PLBART-base\footnote{\href{https://huggingface.co/uclanlp/plbart-base}{huggingface.co/uclanlp/plbart-base}}~\cite{ahmad2021unified} as our backbone models. 

\vspace{-.1em}
\paragraph{Experimental Details.}
All the experiments covered in this paper are conducted on a Linux server with 4 interconnected NVIDIA RTX 3090 GPUs.  
The model is trained with the Adam~\cite{kingma2015adam} optimizer with $\beta_{1} = 0.9$, $\beta_{2} = 0.999$. 

For the source tasks training, we set epoch $e$ as 2 for all cross-language knowledge acquisition. 
And for the cross-task part, 
epoch $e$ is set as 2/6/6 for using code summarization, code understanding~(clone detection and defect detection), and code translation as source tasks respectively.

In order to guarantee a fair comparison, we cannot solely depend on the results stated in prior papers, given that diverse models are evaluated under different settings.
Therefore, we adopt the settings specified in these studies and replicate them on our machines in correspondence with TransCoder.
The details of other settings (for both fine-tuning and TransCoder) are showcased in Appendix~\ref{sec:hyperparams}. 

% Since different PTMs of source code are evaluated on different downstream tasks and dataset, it is impossible to compare them directly based only on the results reported in existing paper.
% source
% cross lang 2
% cross task summarize 2
% cls 6 trans 6
% The experiments cover eight programming languages: Python, Java, C, C\#, Ruby, PHP, JavaScript, and Go.

In the experiments, we set out to answer the research questions (RQs) listed below. We will address each with our results and analysis. 

\begin{itemize}[itemsep=5pt,topsep=0pt,parsep=0pt]
    \item \textbf{RQ1}:~Through learning transferable knowledge, can TransCoder boost mutual enhancement on code downstream tasks?
    \item \textbf{RQ2}:~To what extent can TransCoder help programming languages with small sample sizes to get better results on complex tasks?
    \item \textbf{RQ3}:~For different kinds of target tasks, 
    which source task's knowledge can best promote their performance respectively?
    \item \textbf{RQ4}:~Under low-resource scenarios, to what extent can TransCoder surpass fine-tuning? 
    Additionally, is TransCoder capable of matching the model fine-tuned on the whole dataset?
    % while only using a small percentage of data?
\end{itemize}

\begin{table*}[htb]
\centering
\resizebox{0.7\textwidth}{!}{
\begin{tabular}{lccccccc}
\toprule
Settings & Ruby & JavaScript & Go & Python & Java & PHP & Overall \\
\midrule
\textbf{CodeT5} &  &  &  &  &  &  & \\
Fine-Tuning & 15.24 & 16.21 & 19.53 & 19.90 & 20.34 & 26.12 & 19.56 \\
TransCoder & \textbf{16.88} & \textbf{18.45}  & \textbf{20.40} & \textbf{20.17} & \textbf{21.28} & \textbf{27.28} & \textbf{20.74}\\
\midrule
\textbf{PLBART} &  &  &  &  &  &  & \\
Fine-Tuning & 13.97 & 14.13 & 18.10 & \textbf{19.33} & 18.50 & \textbf{23.56} & 17.93 \\
% ** \\x
TransCoder & \textbf{15.32} & \textbf{15.00}  & \textbf{18.67} & 19.27 & \textbf{19.44} & 23.52 & \textbf{18.54}\\
\bottomrule
    \end{tabular}
    }
    \caption{Comparison between cross-language learning by TransCoder and full fine-tuning, using CodeT5 and PLBART as backbone models.}
    \label{tab:cross-lang-main}
    \vspace{-1em}
\end{table*}

\subsection{Main Results}
\label{section:main-exp}

\paragraph{Effectiveness of Cross-Tasks Scenario.}
Table~\ref{tab:cross-task} shows the main experiment result of cross-task code representation learning by TransCoder. 
Due to the space limit, 
we abbreviated the name of different cross-task scenarios in the form of \texttt{<Source>2<Target>}. 
For instance, 
``CLS2Trans'' refers to using two classification tasks~(clone detection and defect detection) as source tasks and then selecting code translation as the target task. Similarly, ``Trans2Sum'' means employing Java $\rightarrow$ C\# 
 and C\# $\rightarrow$ Java translation as source tasks, and choosing code summarization as the target task.

From these experimental results regarding different source tasks and target tasks, we make the following observations.
1) Learning with TransCoder improves the overall performance of each code downstream task when using different backbones.
2) Regardless of the source tasks involved, target tasks with smaller training sample sizes are most benefited from TransCoder. \textit{e.g.}, there exists a notable improvement in defect detection~(with a modest sample size) in all combinations of backbones and source tasks.
3) These two backbones have comparable parameters, while CodeT5 performs slightly better than PLBART. 
We hold the view that this phenomenon could be attributed to the different pre-training corpus: CodeT5 involves the CodeSearchNet (CSN) dataset, but PLBART does not. 
The universal knowledge is extracted from the bimodal part of the CSN dataset to a large extent, which innately matches the inherent knowledge of CodeT5 and grants it a smoother learning process.

\paragraph{Effectiveness on Cross-Languages Scenario.}
% As is shown in Table~\ref{table:csn-stats}, there is a significant gap between the data volume of Java and Ruby. 
% In the context of code data gathered from platforms such as GitHub, there exists a notable disparity in data volume between Java and Ruby.
Regarding the code data collected from platforms such as GitHub, there exists a significant gap in the volume of data across different PLs. For instance, the data volume for Ruby is substantially smaller, comprising only approximately 15\% of the training sample size available for Java~\footnote{Detailed statistics are available in Appendix\ref{sec:data-stats}.}. 
The latter only has approximately 15\% of the former's training sample size.
Other languages like JavaScript also have a certain degree of data scarcity compared to Java or Python.

Table~\ref{tab:cross-lang-main} demonstrate the main results of cross-language learning by applying TransCoder.
We observe that 
1) The overall performance of summarizing code for different programming languages is enhanced.
2) Different languages get various levels of performance increase with the aid of TransCoder in code summarization. It is worth noticing that
languages with relatively smaller training sample sizes are most benefited from our method. 
Specifically, the performance on summarizing JavaScript code is significantly higher than the baseline, and Ruby also has a notable improvement.
This phenomenon is in line with our expectation that universal code-related knowledge can compensate for the shortcoming of smaller sample sizes.

\subsection{Further Analysis}
\begin{figure}[ht]
    \centering
    \setlength{\abovecaptionskip}{0.cm}
    \includegraphics[width=0.95\linewidth]{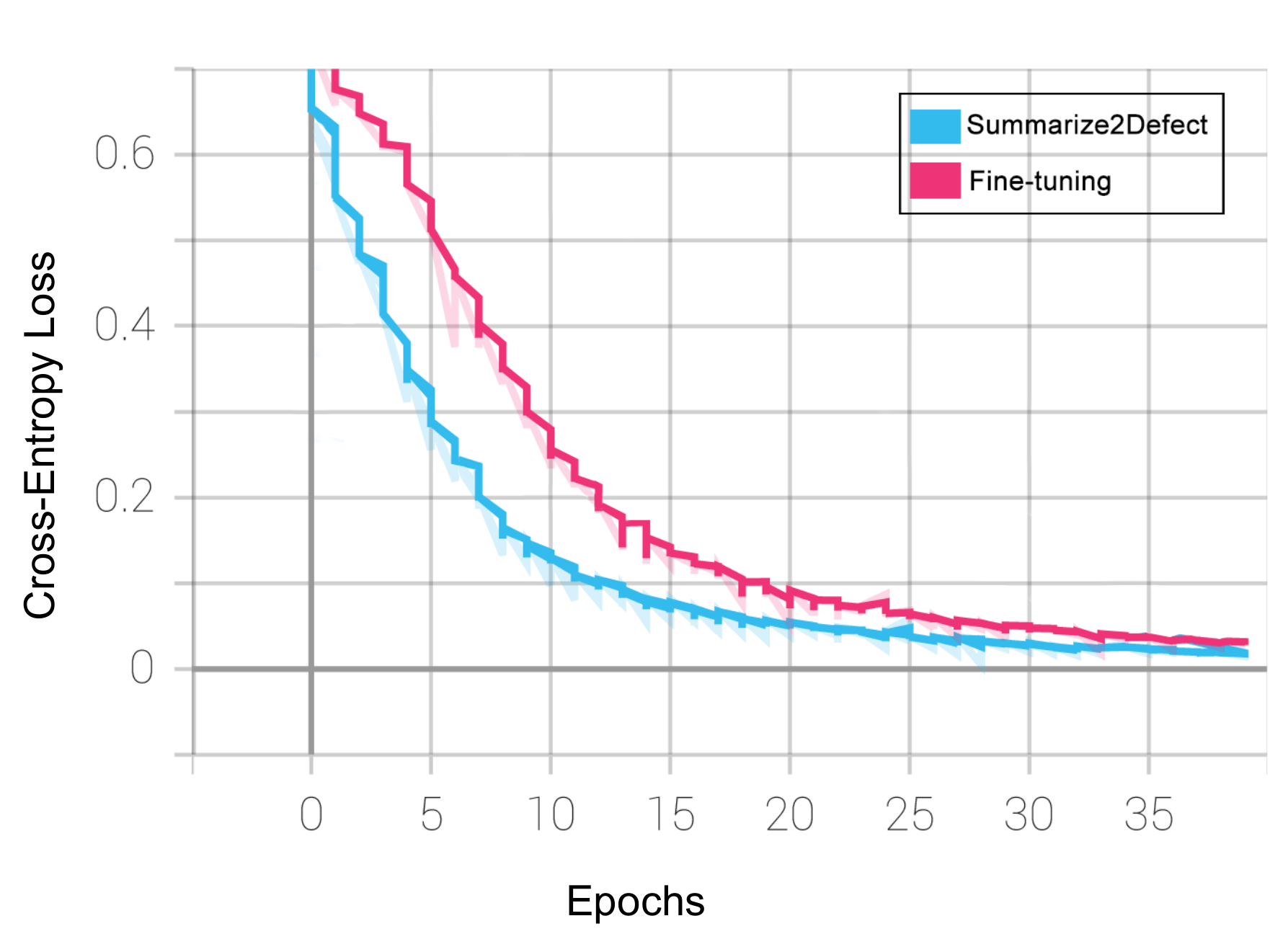}
    \caption{Comparison of code defect detection task between fine-tuning and TransCoder with the knowledge of code summarization (using PLBART backbone).}
    \label{fig:uk-ana}
    \vspace{-1.25em}
\end{figure}

\paragraph{Ablation Study.}
Using the knowledge prefix to absorb and transfer knowledge from different kinds of code-related tasks is one of the key ideas of TransCoder.
Here we conduct an ablation study to understand the effect of using knowledge prefixes. 
We employed a prefix with random knowledge $\theta_0$, \textit{i.e.}, with randomly initialized parameters to verify the effectiveness of the universal knowledge acquisition process on source tasks.
\begin{table}[H]
\centering
\small
\resizebox{0.48\textwidth}{!}{
\begin{tabular}{lcccccc}
\toprule
\multirow{3}{*}{\textbf{Methods}} & \multicolumn{2}{c}{Sum2CLS} & \multicolumn{2}{c}{Trans2CLS} \\
\cmidrule(lr){2-3}\cmidrule(lr){4-5}
 & \makecell[c]{Clone \\ \small{F1}} & \makecell[c]{Defect \\ \small{Acc}} & \makecell[c]{Clone \\ \small{F1}} & \makecell[c]{Defect \\ \small{Acc}} \\
\midrule
\textbf{CodeT5} &  &  &  &  \\
Random Knowl. &  92.38 &  60.76 &  93.68  &  60.29 \\
Universal Knowl. & \textbf{93.70} & \textbf{66.58} & \textbf{95.39} & \textbf{66.36} \\
\midrule
\textbf{PLBART} &  &  &  &  \\
Random Knowl. &  90.96 &  61.02 &  91.15  &  61.64 \\
Universal Knowl. &  \textbf{92.28} & \textbf{64.58} & \textbf{92.91}  & \textbf{64.98} \\

\bottomrule
    \end{tabular}
    }
    \caption{Ablation study of the effectiveness of universal code-related knowledge.
    }
    \label{tab:ablation}
    \vspace{-1.0em}
\end{table}
As is illustrated in Table~\ref{tab:ablation}.
We observe that using the prefix with random knowledge brings worse performance on both two kinds of knowledge transfer, particularly in defect detection.
This further proves the capability of knowledge prefixes to leverage crucial information of code.
\begin{table*}[ht]
\centering
\resizebox{0.7\textwidth}{!}{
\begin{tabular}{lccccccc}
\toprule
Settings & Ruby & JavaScript & Go & Python & Java & PHP & Overall \\
\midrule
\textbf{5\% Data} &  &  &  &  &  &  & \\
Fine-Tuning & 13.96 & 14.68 & 18.04 & 18.30 & 18.74 & 23.42 & 17.86 \\
TransCoder & 14.21 & 15.14 & 19.16 & 19.36 & 18.23 & 23.68 & 18.30 \\
\midrule
\textbf{10\% Data} &  &  &  &  &  &  & \\
% Fine-Tuning & 14.76 & 15.19 & 18.70 & 18.89 & 19.50 & 24.32 & 18.56 \\
Fine-Tuning & 15.22 & 15.12 & 19.06 & 19.20 & 19.32 & 24.95 & 18.81 \\
TransCoder & 16.05 & 16.63 & 20.21 & 20.07 & 20.48 & 26.20 & 19.94 \\
\midrule
\textbf{20\% Data} &  &  &  &  &  &  & \\
% Fine-Tuning & 15.22 & 15.83 & 19.51 & 19.29 & 19.88 & 24.94 & 19.11 \\
Fine-Tuning & 15.23 & 16.01 & 19.44 & 19.91 & 20.38 & 25.51 & 19.41 \\
TransCoder & 16.11 & 17.25 & 20.28 & 20.11 & 20.73 & 26.96 & 20.24 \\
% \midrule
% \textbf{Full Data} &  &  &  &  &  &  & \\
% Fine-Tuning & 15.26 & 16.21 & 19.53 & 19.90 & 20.34 & 26.12 & 19.56 \\
% % ** \\x
% TransCoder & \textbf{16.88} & \textbf{18.45}  & \textbf{20.40} & \textbf{20.17} & \textbf{21.28} & \textbf{27.28} & \textbf{20.74}\\
\bottomrule
    \end{tabular}
    }
    \caption{Varying degrees of low-resource cross-language learning by TransCoder, using code summarization and CodeT5 backbone for evaluation.}
    \label{tab:low-resouce}
    \vspace{-1em}
\end{table*}
\paragraph{Effect of Universal Knowledge.} 
We evaluate the effectiveness of universal code-related knowledge by observing the source task training process. 
Figure~\ref{fig:uk-ana} indicates that with the help of TransCoder, the model can converge faster and reach a new optimum on an unseen code understanding task.
We also prepared another analysis of the code generation task in Appendix~\ref{sec:case-ana}.

% \sqs{need two curves(fine tune \& TransCoder) of experiment}

% sec:order

\subsection{Low-resource Scenarios}
\label{exp:low-resource}

As stated in section~\ref{method:low-resouce}, TransCoder is also capable of conducting code-related downstream tasks under low-resource settings.
To verify our claim, 
we conduct experiments on code summarization\footnote{In the experiment, whenever we specify a language as the target task of code summarization, 
we provide the knowledge of five other languages to TransCoder as source tasks.}, 
which is the most challenging task for code representation learning.
Table~\ref{tab:low-resouce} demonstrates the main results of applying TransCoder to three different scenarios, which only use 5\%, 10\%, and 20\% of the CodeSearchNet~\cite{husain2019codesearchnet} bimodal data\footnote{For a fair comparison, 
we sample the training data from CodeSearchNet at the same specified rate for all languages.} for summarizing code across six programming languages.

It is clear that TransCoder significantly surpasses fine-tuning under the same availability of data in most cases. 
% When tuning on full data, TransCoder can overcome all baselines easily.
Moreover, we only require 10\% of the data to approximate the performance of model tuning on the whole dataset for each language. 
Besides, by utilizing 20\% of the data, the performance of code summarization can significantly surpass fine-tuning the model on the \textbf{whole dataset}.
Additionally, it is worth noticing that the major improvement in code summarization still comes from Ruby and JavaScript. It further confirms that acquiring knowledge from languages with rich corpus can aid subsequent learning from minor datasets.

\section{Conclusion}

In this paper, we proposed \textbf{TransCoder}, 
which is a unified transfer learning framework for both code understanding and generation. 
We draw inspiration from 1) how human beings learn to write codes that the difficulty of learning the next programming language decreases with the aid of previous efforts and experience, and
2) learning multiple code-related tasks will deepen the understanding of the programming domain.
TransCoder employs a two-stage learning strategy that 
1) in the source task training stage, 
our proposed knowledge prefix will absorb universal code-related knowledge through continual learning among various tasks.
2) in the target task specification stage, 
the model will be benefited from the prepared knowledge and utilize it to learn new target tasks.
Extensive experimental results on the benchmark tasks demonstrate that TransCoder significantly enhances the performance of a series of code-related tasks.
% In addition, 
% our approach mitigates the imbalance of training samples in different languages.
Moreover, 
further analysis shows that our method can drive CodePTMs to conduct downstream tasks under low-resource scenarios.
In the future,
we will extend TransCoder to more tasks and incorporate the code structure information into the knowledge acquisition procedure.

\section*{Limitations}
The limitations of our work are shown below:

\begin{itemize}
    \item We primarily rely on the CodeXGLUE Benchmark Dataset~\citep{Lu2021codexglue} to evaluate the effectiveness of our method. 
    We hold the view that incorporating more code corpora with diversity, \textit{e.g.}, BigQuery\footnote{\url{https://console.cloud.google.com/marketplace/details/github/github-repos}}, can further exploit the benefits of TransCoder and extend it to more tasks and languages.
    \item Due to the limited computational resources, 
    our experiments mainly rely on models with encoder-decoder and decoder-only architecture for both code understanding and generation tasks. 
    We leave experiments based on other CodePTMs with encoder-only~\cite{kanade2020learning,feng2020codebert} architecture as future works.
    \item Owing to constraints in data availability, we assess TransCoder’s efficacy under low-resource scenarios by employing a sub-sampling technique on the existing dataset. 
    A more comprehensive evaluation involving less commonly used programming languages in the real world remains a prospect for future endeavors.
\end{itemize}

\section*{Ethical Considerations}

The contribution in this paper, TransCoder, is fully methodological.
Our approach focuses on a new learning framework for code-related downstream applications, 
which encourages mutual reinforcement between tasks and languages. 
It can be applied to a series of scenarios, \textit{e.g.}, low-resource and sample imbalance.
Therefore, this contribution would be beneficial to the NLP community and has no direct negative social/ ethical impacts.

\section*{Acknowledgement}

We thank our anonymous reviewers for their insightful comments and suggestions. This work is supported by Shanghai ``Science and Technology Innovation Action Plan'' Project (No.23511100700).

\bibliography{custom, tacl2021}
\bibliographystyle{acl_natbib}

\appendix

\section{Dataset Statistics}
\label{sec:data-stats}

\begin{table*}[ht]
\begin{center}
\begin{small}
\resizebox{0.7\textwidth}{!}{
\begin{tabular}{lcccccccc}
\toprule
 Methods &  Ruby & JavaScript & Go & Python & Java & PHP & Overall\\
\midrule
\textbf{CodeT5} &  &  &  &  &  &  & \\ 
Fine-Tuning & 15.24 & 16.21 & 19.53 & 19.90 & 20.34 & 26.12 & 19.56 \\
CLS2Sum & 16.62 & 16.54 & 20.67 & 20.07 & 22.41 & 26.03 & 20.39 \\
Trans2Sum & 15.88 & 16.30 & 19.99 & 19.81 & 20.63 & 26.03 & 19.77 \\
\midrule
\textbf{PLBART} &  &  &  &  &  &  & \\ 
Fine-Tuning & 13.97 & 14.13 & 18.10 & 19.33 & 18.50 & 23.56 & 17.93 \\
CLS2Sum & 16.12 & 14.40 & 18.86 & 19.30 & 19.31 & 23.70 & 18.62 \\
Trans2Sum & 14.72 & 14.11 & 18.51 & 19.35 & 19.31 & 23.51 & 18.25 \\
% \bottomrule[.75pt]
\bottomrule
\end{tabular}
}
\caption{Trans2Sum and CLS2Sum}
\label{table:ablation-gat}
% \vspace{-2em}
\end{small}
\end{center}
\end{table*}

\begin{table}[H]
\begin{center}
\resizebox{0.40\textwidth}{!}{
		\begin{tabular}{lccc}
			\toprule
			Language & Training & Dev & Testing \\
			\midrule
			Go&167,288&7,325&8,122\\
			Java&164,923&5,183&10,955\\
			JavaScript&58,025&3,885&3,291\\
			PHP&241,241&12,982&14,014\\
			Python&251,820&13,914&14,918\\
			Ruby&24,927&1,400&1,261\\
			\bottomrule
  \end{tabular}
}
\caption{CodeSearchNet~\cite{husain2019codesearchnet} data statistics for the code summarization task.}
		\label{table:csn-stats}
\end{center}
\end{table}

\begin{table}[H]
\begin{center}
\resizebox{0.48\textwidth}{!}{
		\begin{tabular}{lcccc}
			\toprule
			Dataset & Language & Training & Dev & Testing \\
			\midrule
			CodeTrans & Java - C\# & 10,300 & 500 & 1,000 \\
			\bottomrule
		\end{tabular}
		}
		\caption{CodeTrans~\cite{nguyen2015divide} datasets statistics for code translation task.}
		\label{table:code-trans}
\end{center}
\end{table}

\begin{table}[H]
\begin{center}
\resizebox{0.48\textwidth}{!}{
		\begin{tabular}{lcccc}
			\toprule
			Dataset & Language & Training & Dev & Testing \\
			\midrule
			BigCloneBench & Java & 900K & 416K & 416K \\
			Devign & C & 21K & 2.7K & 2.7K \\
			\bottomrule
		\end{tabular}
		}
\caption{BigCloneBench~\cite{svajlenko2014towards} and Devign~\cite{zhou2019devign} datasets statistics for Clone detection and Defect Detection tasks.}
    \label{table:clone-defect-stats}
\end{center}
\end{table}

\section{Case Analysis}
\label{sec:case-ana}

\begin{figure}[htb]
    \centering
    \includegraphics[width=0.85\linewidth]{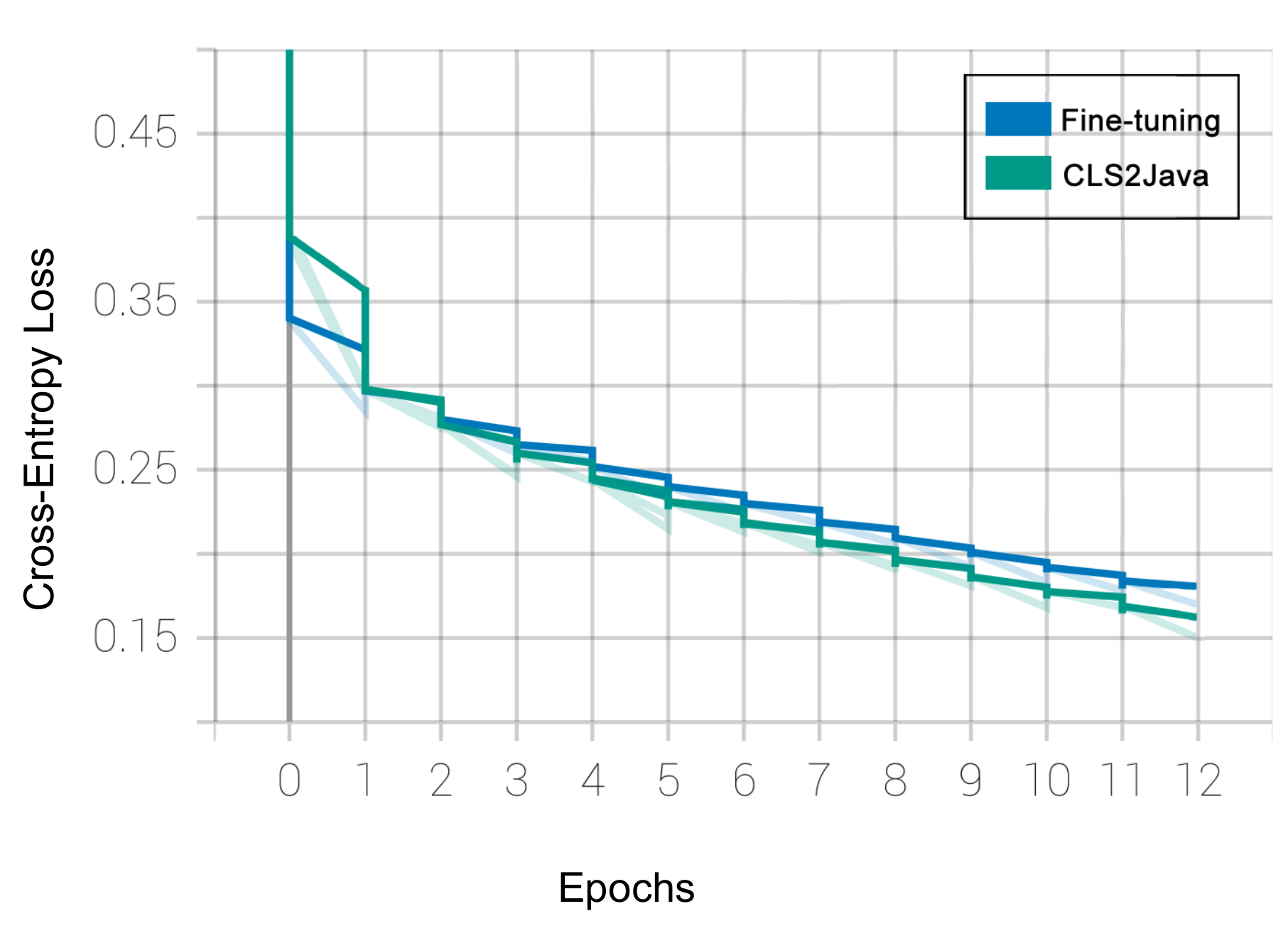}
    \caption{Comparison between TransCoder (with code understanding knowledge) and fine-tuning on summarizing Java code
    based on CodeT5 backbone.}
    \label{fig:uk-ana-codet5}
\end{figure}

Figure~\ref{fig:uk-ana-codet5} demonstrates the effectiveness of universal code-related knowledge when using CodeT5 as the backbone model.

It is clear that, through involving TransCoder, the loss function can be further optimized on the training set.

\section{Details of Code Summarization}

Due to the space limit, 
we report the overall performance for code summarization in section~\ref{section:main-exp}. The details of summarizing code snippets of six different programming languages based on two backbone CodePTMs are given in Table~\ref{table:ablation-gat}.

\section{Hyperparameters Settings}
\label{sec:hyperparams}

The hyperparameters for model fine-tuning (baselines) and TransCoder are listed in~\ref{tab:hparam-ft} and Table~\ref{tab:hparam}.
% The hyperparameters of TransCoder are listed in 

\begin{table}[htb]
\centering
\resizebox{0.4\textwidth}{!}{
% \centering
\begin{tabular}{lc}
\toprule
\textbf{Hyperparameter} & \textbf{value} \\
\midrule
Batch Size & {8,16} \\
Learning Rate & \{8e-6, 2e-5\} \\
% Adam Epsilon &  1e-8 \\
Max Source Length & \{256, 320, 512\} \\
Max Target Length & \{3, 128, 256, 512\} \\
Epoch & \{2, 30, 50, 100\}\\
\bottomrule
\end{tabular}
}
\caption{
Hyperparameters for fine-tuning
}
\label{tab:hparam-ft}
\end{table}

\begin{center}
\begin{table}[htbp]
\centering
\resizebox{0.4\textwidth}{!}{
% \centering
\begin{tabular}{lc}
\toprule
\textbf{Hyperparameter} & \textbf{value} \\
\midrule
Batch Size & {8,16,32} \\
Learning Rate & \{2e-5, 1e-4, 5e-4\} \\
% Learning Rate & \{8e-6, 1e-5, 2e-5, 5e-5\} \\
% Adam Epsilon &  1e-8 \\
% Max Source Length & \{130, 240, 256, 320, 512\} \\
% Max Target Length & \{3, 120, 150, 240, 256, 512\} \\
Prefix Length & 32\\
Source Train Epoch & \{2, 4, 6\}\\
Target Train Epoch & \{2, 30, 50, 100\}\\
Smoothing Factor $\delta$ & \{0.5, 1\} \\
\bottomrule
\end{tabular}
}
\caption{
Hyperparameters for TransCoder
}
\label{tab:hparam}
\end{table}
\end{center}

\section{Experiments on Tasks Training Orders}
\label{sec:order}

\begin{figure}[htbp]
\setlength{\abovecaptionskip}{0.cm}
    \centering
    \subfigure[ \uppercase\expandafter{\romannumeral1}]{
        \includegraphics[width=3.2cm]{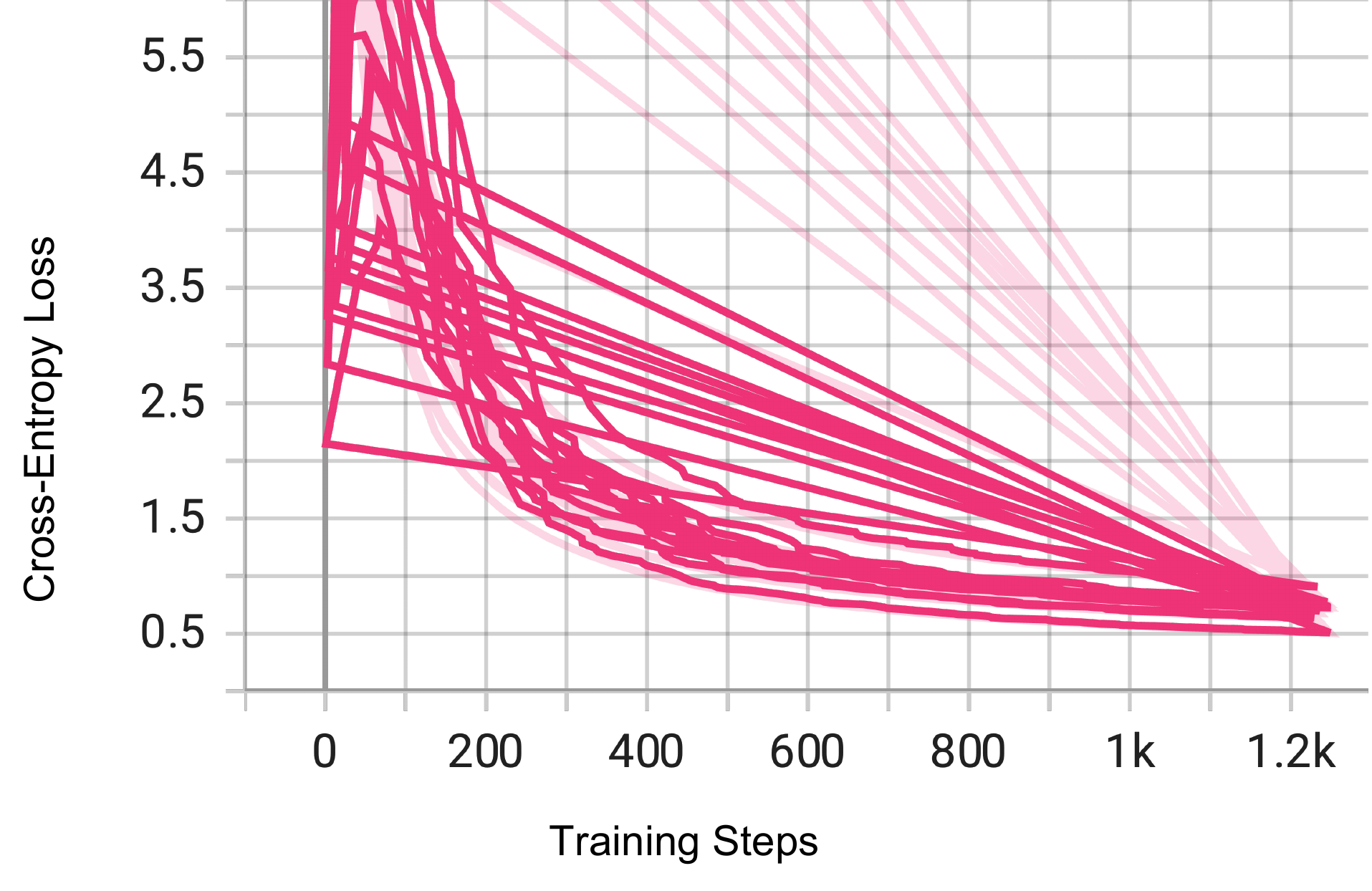}
    }
    \subfigure[
    \uppercase\expandafter{\romannumeral2}
    ]{
	\includegraphics[width=3.2cm]{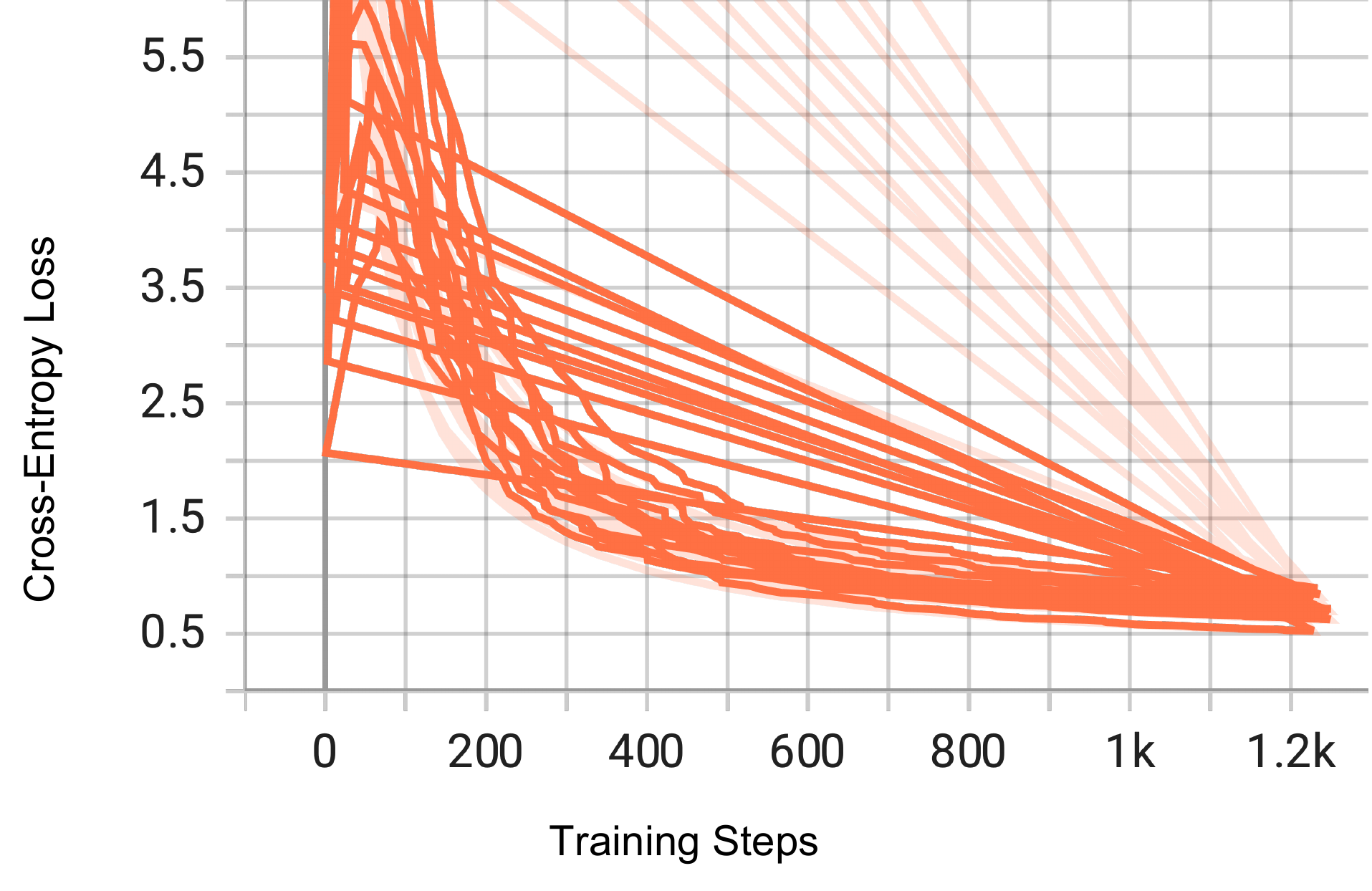}
    }
    \subfigure[
    \uppercase\expandafter{\romannumeral3}
    ]{
    	\includegraphics[width=3.2cm]{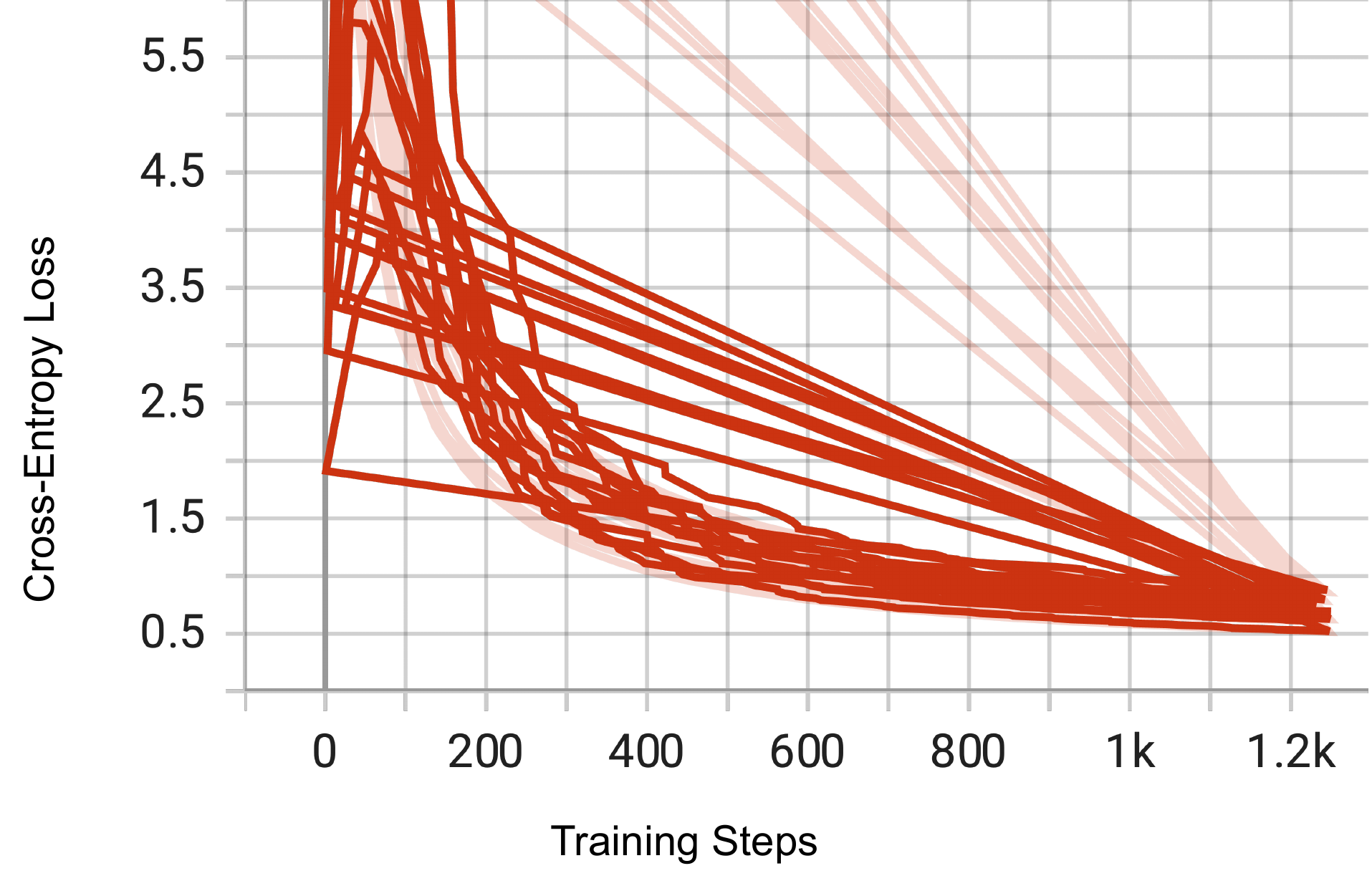}
    }
    \subfigure[
    \uppercase\expandafter{\romannumeral4}
    ]{
	\includegraphics[width=3.2cm]{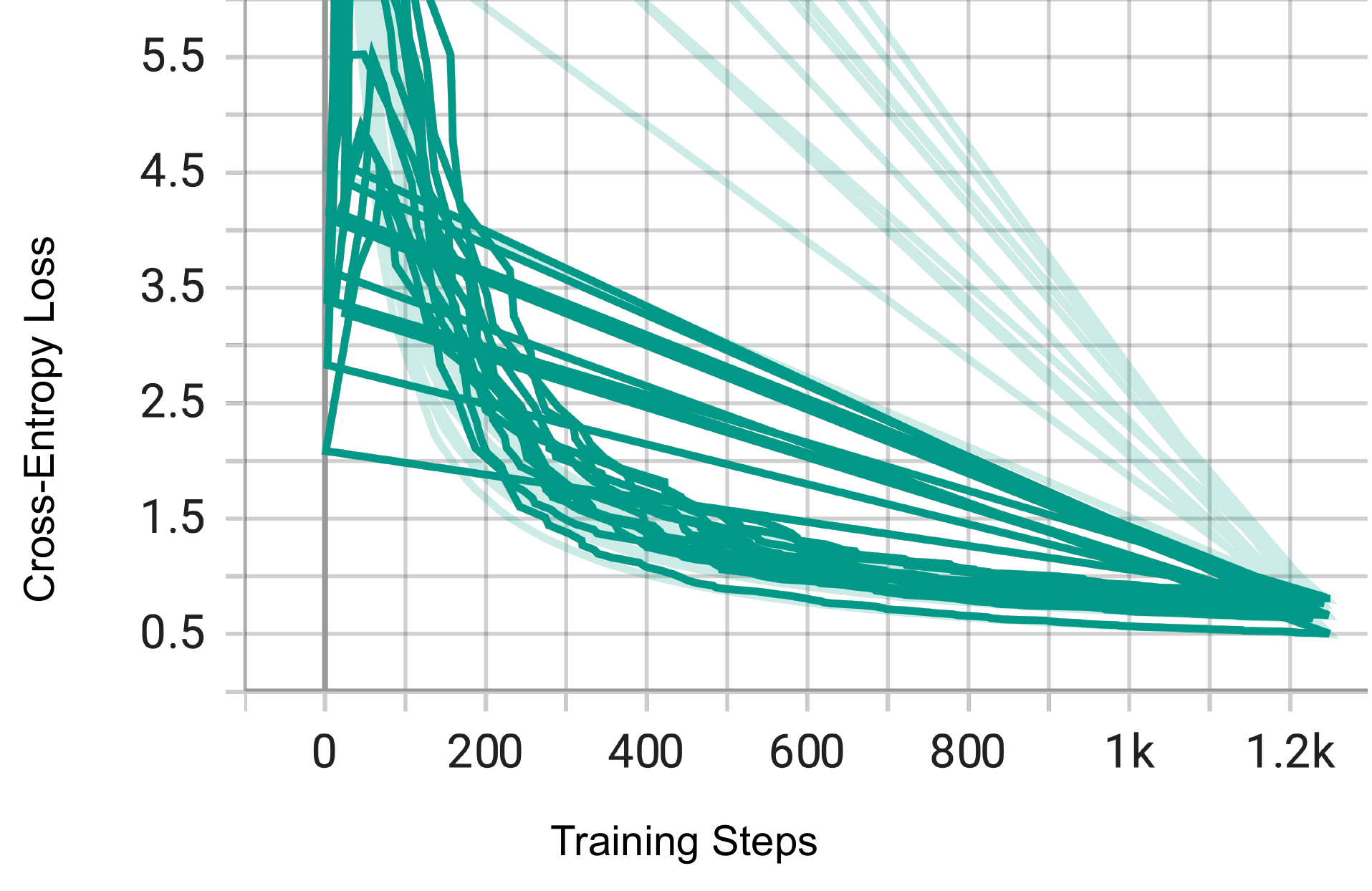}
    }
    \caption{Employing different source tasks training order, refer Table~\ref{tab:task-order} for details.}

    \label{fig:source-order}
\end{figure}

% \paragraph{Order of Training Tasks.}
In order to gauge the stability of TransCoder, we perform experiments on the training sequence of source tasks.
% We also conduct experiments on source tasks training order to test the stability of TransCoder.
We train our universal knowledge prefix for cross-language learning in various orders.
The detailed results are shown in Figure~\ref{fig:source-order} and Table~\ref{tab:task-order}.
There exists no significant difference in training,
and problems like "catastrophic forgetting" do not occur.
This also verifies the effectiveness of our adaptive sampling strategy that prevents the knowledge prefix from being over-reliant on certain tasks.

The training curves derived from different source task orders are depicted in Figure~\ref{fig:source-order}. 
The number range from 1 to 6 in Table~\ref{tab:task-order} indicates the exact order when the task is selected.

\begin{table}[H]
\centering
\resizebox{0.45\textwidth}{!}{
\begin{tabular}{ccccccc}
\toprule
 Training Order & Ruby  & JavaScript & Go & Python & Java & PHP  \\
\midrule
  \uppercase\expandafter{\romannumeral1} & 3 & 5 & 6 & 4 & 1 & 2  \\
  \uppercase\expandafter{\romannumeral2} & 4 & 6 & 5 & 3 & 2 & 1  \\
  \uppercase\expandafter{\romannumeral3} & 1 & 2 & 3 & 6 & 4 & 5 \\
  \uppercase\expandafter{\romannumeral4} & 3 & 4 & 5 & 1 & 6 & 2  \\
\bottomrule
\end{tabular}
    }
    \caption{Comparison between employing different source tasks training order by cross-language learning of TransCoder.  
    % The BLEU score of summarizing Ruby code is reported, and the number range from 1 to 5 indicates the training order among five languages.
    }
    \label{tab:task-order}
\end{table}

\end{document}